\documentclass[conference]{IEEEtran}
\usepackage[latin9]{inputenc}
\usepackage{array}
\usepackage{float}
\usepackage{booktabs}
\usepackage{amsmath}
\usepackage{amsthm}
\usepackage{amssymb}
\usepackage{graphicx}

\makeatletter

\providecommand{\tabularnewline}{\\}
\floatstyle{ruled}
\newfloat{algorithm}{tbp}{loa}
\providecommand{\algorithmname}{Algorithm}
\floatname{algorithm}{\protect\algorithmname}

\theoremstyle{plain}
\newtheorem{thm}{\protect\theoremname}

\usepackage{amsfonts}
\usepackage{cite}
\usepackage{algorithm}
\usepackage{algpseudocode}
\algnewcommand{\Linecomment}[1]{\Statex \(\triangleright\) #1}
\algrenewcommand\algorithmicrequire{\textbf{Input:}}
\algrenewcommand\algorithmicensure{\textbf{Output:}}
\usepackage[justification=centering]{caption}

\usepackage{array}
\usepackage{stfloats}
\usepackage{url}
\usepackage{verbatim}
\def\BibTeX{{\rm B\kern-.05em{\sc i\kern-.025em b}\kern-.08em
    T\kern-.1667em\lower.7ex\hbox{E}\kern-.125emX}}
\usepackage{balance}

\newtheorem{lemma}{Lemma}

\makeatother

\providecommand{\theoremname}{Theorem}

\begin{document}
\title{Efficient Coflow Scheduling in Hybrid-Switched Data Center Networks}
\author{{\normalsize{}Xin Wang$^{\dagger}$, Hong Shen$^{\ast,\dagger}$,
Hui Tian$^{\ddagger}$}\\
{\normalsize{}}\\
{\normalsize{}$^{\dagger}$School of Computer Science and Engineering,
Sun Yat-sen University, China}\\
{\normalsize{}$^{\ast}$School of Applied Sciences, Macao Polytechnic
University, Macao SAR, China}\\
{\normalsize{}$^{\ddagger}$School of Information and Commnucation
Technology, Griffith University, Australia}}
\maketitle
\begin{abstract}
To improve the application-level communication performance, scheduling
of coflows, a collection of parallel flows sharing the same objective,
is prevalent in modern data center networks (DCNs). Meanwhile, a hybrid-switched
DCN design combining optical circuit switches (OCS) and electrical
packet switches (EPS) for transmitting high-volume and low-volume
traffic separately has recently received considerable research attention.
Efficient scheduling of coflows on hybrid network links is crucial
for reducing the overall communication time. However, because of the
reconfiguration delay in the circuit switch due to the ultra-high
transmission rate and the limitation of bandwidth in the packet switch,
coflow scheduling becomes increasingly challenging. The existing coflow
scheduling algorithms in hybrid-switched DCNs are all heuristic and
provide no performance guarantees. In this work, we propose an approximation
algorithm with the worst-case performance guarantee of $O(\tau)$,
where $\tau$ is a factor related to system parameters and demand
characteristics, for single coflow scheduling in hybrid-switched DCNs
to minimize the coflow completion time (CCT). Extensive simulations
based on Facebook data traces show that our algorithm outperforms
the state-of-the-art schemes Solstice by 1.08$\times$ and Reco-Sin
by 1.42$\times$ in terms of minimizing CCT.
\end{abstract}


\section{Introduction}

\begin{figure}[tbh]
\centering \includegraphics[width=0.23\textwidth,height=2.3cm]{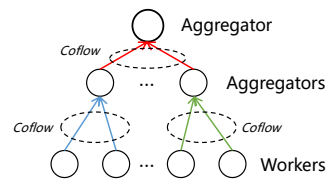}
\caption{Coflow Abstraction in Partition-Aggregate Pattern}
\label{fig:partition}
\end{figure}

As the emergence of data center networks (DCNs), various data parallel
frameworks such as MapReduce \cite{mapreduce}, Spark \cite{spark}
and Dryad \cite{dryad} are gaining increasing popularity. The execution
process of typical data-parallel applications usually consists of
multiple consecutive stages. Each stage is dependent on a collection
of parallel flows, termed $\textit{coflow}$ \cite{networking}, and
the next stage cannot begin until all flows in the current stage have
completed their transmission. Hence, application-level performance
largely depends on coflow completion time (CCT), i.e., the completion
time of the slowest flow within a coflow, and minimizing the CCT becomes
an interesting problem of great significance for improving application-level
performance. It should be noted that applying the traditional network
metrics, such as minimizing the flow completion time (FCT), is unable
to minimize the CCT and therefore ineffective for improving the communication
performance of coflows in the application. Coflow is a higher-level
networking abstraction that captures a range of communication patterns
observed in cluster computing applications, such as Partition-Aggregate,
Bulk Synchronous Parallel, and Shuffle. As shown in Fig. \ref{fig:partition},
the coflow abstraction accurately reflects the communication pattern
during partition-aggregate communication.

Based on the coflow abstraction, many coflow scheduling algorithms
\cite{literature4,literature6,literature7,literature9,literature10,literature11}
have been designed to improve the traffic transmission efficiency
in DCNs supported by conventional Electronic Packet Switches (EPS).
Varys \cite{literature6} proposed the smallest-effective-bottleneck-first
(SEBF) and minimum-allocation-for-desired-duration (MADD) heuristic
algorithms that greedily schedule coflows based on the bottleneck
completion time of coflow to minimize the CCT. Barrat \cite{literature7}
and Stream \cite{literature8} both focused on decentralized coflow
scheduling. Aalo \cite{Aalo} utilizes Discretized Coflow-Aware Least-Attained
Service (D-CLAS) to schedule coflows without prior knowledge of coflows.
Additionally, several theoretical studies \cite{literature9,literature10,literature11}
have been proposed with the aim to minimize the total weighted CCT.
While packet switches have advantages for flow transmission, such
as the ability to make forwarding decisions at the packet level, their
bandwidth grows too slowly to meet the demands of modern DCNs.

As a result, optical circuit switches (OCS) have become more common
in contemporary DCNs to meet network bandwidth demands. Compared to
traditional EPS, OCS offers much better data transfer rates and lower
power consumption. However, the transmission mode of OCS limits each
ingress or egress port only to establish one circuit at a time, called
the \textit{port constraint}. In addition, each circuit reconfiguration
in OCS incurs a fixed time delay $\delta$ (i.e., a reconfiguration
delay), typically between a few hundred microseconds and a few tens
of milliseconds. Therefore, the above EPS-based scheduling algorithms
cannot be directly applied to OCS, which would violate the \textit{port
constraint}, as they all provide bandwidth sharing, i.e., an ingress
(egress) port may be connected to multiple egress (ingress) ports
simultaneously. Several EPS-based flow scheduling algorithms \cite{sincronia,animproved}
do not share bandwidth and are therefore suitable for OCS. However,
these approaches may require frequent circuit reconfigurations. Hence,
coflow scheduling in OCS still faces several difficulties.

The literature on coflow scheduling in optical circuit switches (OCS)
is still limited. To our knowledge, Sunflow \cite{sunflow} is the
first study to consider both OCS and coflow characteristics. OMCO
\cite{omco} proposes a heuristic algorithm for scheduling coflow
in OCS under online scenarios, but does not guarantee performance.
Reco-Sin \cite{regularization} is the first constant approximation
algorithm for single coflow scheduling in OCS, with an approximation
ratio of 2. In addition, researchers have proposed hybrid-switched
DCNs which combine high-speed OCS with traditional EPS to offer higher
throughput at a reduced cost, such as Helios \cite{helios} and c-Through
\cite{c-through}. However, these methods aim to minimize the flow
completion time (FCT) rather than coflow completion time (CCT). Recently,
Liu et al. \cite{solstice} have provided an exciting flow scheduling
heuristic algorithm (called Solstice) that minimizes the maximum flow
completion time in a hybrid network. The algorithm effectively improves
circuit utilization and reduces the number of configurations.

To efficiently schedule coflows in a hybrid network, we need to determine:
(1) a set of circuit configurations in the OCS, i.e., what ports are
connected and the connection durations; and (2) which traffic should
be allocated to the packet switch. Considering the delay time between
configurations, it is necessary to reduce the frequency of reconfigurations
in OCS. Meanwhile, we need to coordinate traffic demand allocation
among different switches in order to achieve high link utilization
and reduce the total completion time (i.e., CCT) in a hybrid-switched
network. In this paper, we propose a new and effective operation,\textit{
}called \textit{Migrating}, which distributes as much traffic load
to the packet switch and balances traffic across the packet and circuit
switch. By ensuring simultaneous completion of transmission in both
switches, \textit{Migrating }can maximize link utilization over the
packet switch and further reduce the CCT. This paper investigates
single coflow scheduling problem in hybrid-switched DCNs, aiming at
minimizing the CCT, while providing detailed theoretical analysis
and proof of approximation ratio. We can summarize the contribution
of this work as follows:
\begin{itemize}
\item We propose a new and effective operation, called \textit{Migrating},
which can further reduce the CCT and optimizes the system performance.
\item We further propose an efficient coflow scheduling algorithm to minimize
the CCT and show that it achieves a performance guarantee (approximation
ratio to the optimal solution) of $O(\tau)$, where $\tau$ is the
demand characteristic. To our knowledge, this is the first approximation
algorithm for coflow scheduling in hybrid-switched DCNs, and hence
fills a research gap.
\item We evaluate our method's performance using real-life traces from Facebook.
Simulation results demonstrate that our method outperforms the state-of-the-art
schemes regarding the reduced number of reconfigurations and faster
transmission of a single coflow.
\end{itemize}
The rest of the paper is organized as follows. Section II defines
the system model and formulates the problem. Section III describes
our proposed coflow scheduling algorithm in hybrid networks. Section
IV provides specific theoretical analysis and approximation ratio
proof. Section V presents the experimental results of our algorithm
and performance comparison with the state-of-the-art works. Finally,
Section VI concludes the paper.

\section{Model and Problem Formulation}

This section presents the system model and the formal definition of
the coflow scheduling problem in a hybrid network.

\subsection{System Model}

\begin{figure}[h]
\centering \includegraphics[width=0.4\textwidth,height=3.5cm]{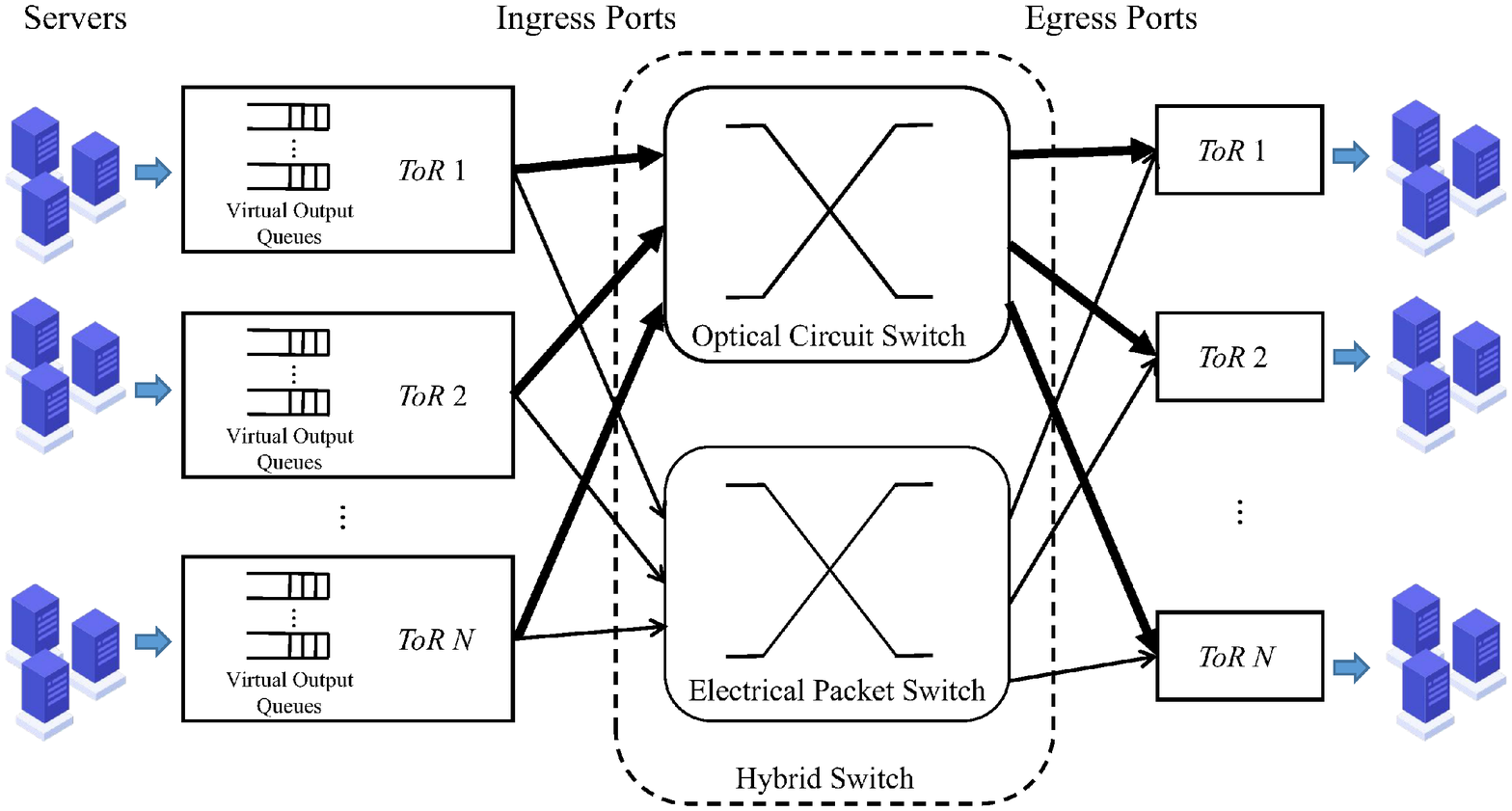}
\caption{A Hybrid Switch Architecture}
\label{fig:hybrid}
\end{figure}

\textit{Network Model}: In data center networks (DCNs), two types
of switches are typically used: optical circuit switches (OCS) and
electrical packet switches (EPS). As shown in Fig. \ref{fig:hybrid},
the DCN is modeled as a non-blocking hybrid circuit/packet switch
with $N$ ingress ports and $N$ egress ports, where each ingress/egress
port is connected to both a circuit switch and a packet switch. In
many cases, these ports are connected to Top-of-Rack (ToR) switches,
with each ToR switch connecting to a group of machines. At the sender
machines, flows are temporarily buffered, aggregated and organized
into virtual output queues (VOQs) for each ingress port. Circuit switches
can only handle one VOQ at a time per ingress port, while packet switches
can handle multiple VOQs, simultaneously.

\textit{Optical Circuit Switch}: Compared with EPS, OCS is capable
of higher data transfer rates and lower power consumption, which makes
it suitable for high-bandwidth applications. OCS needs to establish
a circuit between ingress and egress ports so that data can be transferred
between them. However, OCS has a port connection restriction known
as \textit{port constraint}. Specifically, at any given time, each
ingress (egress) port can only have one circuit connection to one
egress (ingress) port. In addition, OCS requires the reconfiguration
of a new circuit, resulting in a delay in the reconfiguration process
of up to ten microseconds. Until the reconfiguration process is complete,
all transmissions in the OCS may be halted, which is called \textit{all-stop}
circuit switch model and is widely used in existing works \cite{article7,c-through,article16}.
In addition, Sunflow \cite{sunflow} adopts a \textit{not-all-stop}
circuit establishment model, allowing the transmission to continue
on unchanged circuits during reconfiguration and stop only on affected
ports. However, implementing a pure \textit{not-all-stop} OCS remains
challenging due to the immaturity of current manufacturing technology
\cite{regularization}. Therefore, in our work, we also adopt the\textit{
all-stop} model.

\textit{Electrical Packet Switch}:\textit{ }Compared with OCS, a significant
advantage of EPS is its ability to provide more flexible network connectivity
without the \textit{port constraint} of OCS. In EPS, each ingress
or egress port can connect to multiple egress or ingress ports at
the same time, enabling port-sharing and providing greater flexibility
in network configuration. However, this flexibility requires careful
management of available bandwidth to ensure that the network can satisfy
the bandwidth requirements of all simultaneous connections, which
is known as \textit{bandwidth constraint}.

\subsection{Problem Formulation}

We now formally define the coflow scheduling problem for single coflow
in a hybrid network fabric. The main mathematical notations used are
listed in Table I.

\begin{table}[h]
{}{}\vspace*{0.8\baselineskip}
 \caption{Notations}
\label{tab:notation1} %
\begin{tabular}{lp{0.8\linewidth}}
\toprule 
{}{}Symbol & {}{}Definition\tabularnewline
\midrule 
{}{}$N$ & {}{}The number of hybrid switch ports\tabularnewline
{}{}$\delta$ & {}{}The fixed reconfiguration delay time\tabularnewline
{}{}$r_{c}$ & {}{}The circuit link rates\tabularnewline
{}{}$r_{p}$ & {}{}The packet link rates\tabularnewline
{}{}$D$ & {}{}The input demand matrix ($N\times N$)\tabularnewline
{}{}$\rho$ & {}{}The maximum value of the sum of each row and column of $D$,
that is the diameter of $D$\tabularnewline
{}{}$\tau$ & {}{}The maximum number of non-zero elements of each row or column
of $D$\tabularnewline
{}{}$E$ & {}{}The residual demand sent to packet switch ($N\times N$)\tabularnewline
{}{}$L$ & {}{}The number of configurations\tabularnewline
{}{}$P_{l}$ & {}{}The $l$-th circuit switch configuration (permutation) matrix
($N\times N$)\tabularnewline
{}{}$t_{l}$ & {}{}The time duration of $P_{l}$\tabularnewline
{}{}$T$ & {}{}The total completion time of traffic demand\tabularnewline
{}{}$T_{trans}$ & {}{}The total transmission time of traffic demand\tabularnewline
{}{}$T_{conf}$ & {}{}The total configuration delay time of traffic demand\tabularnewline
\bottomrule
\end{tabular}{}{}
\end{table}

We denote $T_{c}$ as the total completion time (i.e., CCT) of coflow
$c$, given by $T_{c}=\max(T_{c}^{O},T_{c}^{E})$, where $T_{c}^{O}$
is the total completion time of the traffic demand transmitted through
the circuit switch and $T_{c}^{E}$ is the total completion time of
the residual traffic demand transmitted through the packet switch.
In hybrid networks, the circuit switch is primarily used for transmitting
most of the traffic, and it is commonly assumed that $T_{c}^{E}\leq T_{c}^{O}$,
implying that $T_{c}=T_{c}^{O}$. Therefore, unless otherwise specified,
the total completion time of the coflow specifically refers to $T_{c}^{O}$,
which includes the total transmission time and the total configuration
delay time on the circuit switch.

\textit{Problem Definition} (\textit{Single Coflow Scheduling}):\textit{
}Given\textit{ }a demand matrix $D$ of coflow, we want to compute
a feasible coflow schedule in an $N\times N$ non-blocking hybrid
network to reduce the number of configurations and minimize the coflow
completion time (CCT).

\textit{Input (traffic demand): }The communication requirements of
a coflow can be represented by an $N\times N$ demand matrix $D$.
Each element $d_{i,j}\in D$ corresponds to the amount of data that
flow $f_{i,j}$ needs to transmit from ingress port $i$ to egress
port $j$, where $1\leq i,j\leq N$.

\textit{Output (Scheduling)}:\textit{ }The output of the scheduling
process has two main components. The first component of the output
is a circuit switch schedule, denoted as $(M,P_{l},t_{l})$, which
consists of a set of configurations $\left\{ P_{1},P_{2},\ldots,P_{L}\right\} $
and the corresponding durations $\left\{ t_{1},t_{2},\ldots,t_{L}\right\} $.
In the circuit switch, each configuration $P_{l}(1\leq l\leq L)$
encodes the connectivity of ports as a $N\times N$ binary matrix.
Specifically, $P_{l}^{i,j}$ is set to 1 if port $i$ is allowed to
send data to port $j$ during this configuration. All $P_{l}$ are
permutation matrices, that is, they have exactly one 1 in each row
and column due to the fact that the circuit switch establishes a one-to-one
connection between each sender and receiver. Each configuration $P_{l}$
also has a duration $t_{l}$ that specifies how long the circuit switch
should stay in the specific configuration. The second component of
the output is the residual demand, denoted as $E$, which is also
an $N\times N$ matrix. The elements $E_{i,j}$ in $E$ represent
the demand $D_{i,j}$ that is routed from port $i$ to port $j$ via
the packet switch.

\textit{Objective}:\textit{ }In a hybrid network, our scheduling objective
is to minimize the total completion time for scheduling the entire
traffic demand $D$. To achieve this goal, we need to effectively
reduce the frequency of reconfigurations in OCS, and allocate the
traffic demand reasonably while allowing packet switches to carry
as much traffic load as possible.

The following is a formal definition of our scheduling objective,
as well as two constraints related to demand satisfaction and packet
switch capacity.\textit{ }More specifically, we define the total completion
time as the sum of the time required to transfer the demand on the
circuit switch (i.e., the total transmission time, $T_{trans}$) and
the time required to wait for reconfiguration delays when switching
between configurations (i.e., the total configuration time, $T_{conf}$).
Our objective is therefore

\begin{equation}
\begin{aligned}\label{eq:totaltime}\min CCT=\min\left(T_{trans}+T_{comf}\right)\\
=\min\left(\left(\sum_{l=1}^{L}t_{l}\right)+L\delta\right).
\end{aligned}
\end{equation}

$\textbf{Demand satisfaction constraint:}$ It is required that the
sum of data transferred via the packet switch and data transferred
via the circuit switch must be greater than or equal to the demand
between the source and destination for each ingress-engress pair in
traffic demand $D$, i.e, covering the whole traffic demand $D$:

\begin{equation}
\begin{aligned}\label{eq:demand}E+\sum_{l=1}^{L}r_{c}t_{l}P_{l}\geq D.\end{aligned}
\end{equation}

$\textbf{Packet switch capacity constraint:}$ The data capacity that
can be carried by a packet switch is constrained by the time spent
on the circuit switch, as both must occur concurrently. Therefore,
for each ingress $i$ or engress $j$ in the packet switch, the allowable
amount of data is limited:

\begin{equation}
\begin{aligned}\label{eq:capacity}\sum_{j=1}^{N}E_{i,j}\leq r_{p}T,\forall i\in\{1,2,\ldots,N\},\\
\sum_{i=1}^{N}E_{i,j}\leq r_{p}T,\forall j\in\{1,2,\ldots,N\}.
\end{aligned}
\end{equation}

Since the coflow scheduling problem is known to be NP-hard \cite{sunflow,animproved},
we will propose an efficient approximation algorithm for coflow scheduling
in hybrid-switched DCNs in the next section.

\section{The Algorithm}

Birkoff-von Neumann (BvN) \cite{BvN} decomposition is a traditional
and classical method used to schedule coflow in optical circuit switches
(OCS). However, in hybrid network environments, the basic BvN method
has two main limitations: (1) it only considers the circuit switch
without utilizing the necessary packet switch, and (2) it does not
address the problem of minimizing the number of configurations of
the OCS, and thus may result in possible delays in reconfigurations
\cite{solstice}. To overcome these limitations, we propose an efficient
coflow scheduling algorithm specifically for hybrid networks that
has a provable performance guarantee.

\subsection{Birkoff-von Neumann Decomposition}

The Birkhoff-von Neumann (BvN) decomposition (as shown in Algorithm
\ref{alg:alg1}) requires an input matrix $D$ of size $N\times N$,
with each element being non-negative and the sum of each row and column
equaling a constant value $K$, that is known as a $K$-bistochastic
matrix. According to the BvN theorem, any $K$-bistochastic matrix
can be decomposed into a set of up to $N^{2}$ permutation matrices,
whose non-negative sum of durations is $K$. However, finding the
optimal BvN decomposition with the least permutation matrices is an
NP-hard problem \cite{dufosse2016notes}. In fact, the demand matrix
in practical applications may not naturally be $K$-bistochastic,
but by adding artificial demands, a pre-processing method called \textit{Stuffing
}\cite{solstice,regularization} can transform it into one.

\begin{algorithm}
\global\long\def\thealgorithm{1}%
s\caption{BvN \cite{BvN} Decomposition Algorithm}
{\begin{algorithmic}[1] \Require $K$-bistochastic traffic matrix
$D$ ($N\times N$); the circuit link rates $r_{c}$\Ensure a circuit
switch schedule: $(M,P_{l},t_{l})$ \State$l\leftarrow1$\While{$D>0$}\State$B\leftarrow$
BinaryMatrix ($D$) \State Interpret $B$ as a bipartite graph of
senders to receivers \State Calculate a perfect matching $P_{l}$
of $B$ \State Interpret $P_{l}$ as a permutation matrix \State$t_{l}\leftarrow\min\left\{ D_{i,j}\mid P_{l}^{i,j}=1\right\} /r_{c}$
\State $D\leftarrow D-r_{c}t_{l}P_{l}$ \State$l\leftarrow l+1$\EndWhile\State
$L\leftarrow l-1$\end{algorithmic} \label{alg:alg1} }
\end{algorithm}

In practice, due to preemption, the BvN decomposition often generates
schedules with many configurations when the demand matrix has a large
ratio between its maximum and minimum non-zero elements (i.e., the
matrix is highly skewed), resulting in extensive reconfigurations.
The durations of these configurations are usually quite short (e.g.,
on the order of the reconfiguration delay $\delta$), resulting in
lower overall efficiency. The problem, however, is that the BvN decomposition
must provide service for the whole traffic demand, including configurations
with shorter durations that have lower efficiency. Therefore, in a
hybrid network, if some demands can be transferred by a packet switch,
the scheduling algorithm can focus on finding configurations that
can last longer, resulting in higher efficiency.

\subsection{Approximation Algorithm}

The main challenge for coflow scheduling in hybrid networks is how
to effectively reduce the frequency of reconfigurations in OCS while
allowing packet switches to carry as much traffic load as possible
thus minimizing the total completion time (i.e., CCT). We propose
an efficient approximate algorithm for coflow scheduling in hybrid
networks, as depicted in Algorithm \ref{alg:alg2}.

\begin{algorithm}
\global\long\def\thealgorithm{2}%
\caption{Coflow Scheduling in Hybrid Networks}
{\begin{algorithmic}[1] \Require the traffic demand $D\left(N\times N\right)$;
the circuit reconfiguration delay time $\delta$; the circuit and
packet link rates $r_{c}$ and $r_{p}$ \Ensure $L$ circuit configurations
and corresponding durations: ${P_{l}}$, ${t_{l}}$; the residual
demand sent to packet switch $E\left(N\times N\right)$ \State$D^{\prime}\leftarrow$run
\textit{Regularization} and \textit{Stuffing} on $D$ with $\delta$
\State$T^{\prime}\leftarrow0$\State$\gamma\leftarrow\text{ largest power of }2\text{ smaller than }\max\left(D^{\prime}\right)$\State$l\leftarrow1$\While{$r_{p}T^{\prime}<\rho_{E^{\prime}}$}\State$P_{l}\leftarrow$
\textit{Slicing} $\left(D^{\prime},\gamma\right)$\If{$P_{l}\neq NULL$
} \State $t_{l}\leftarrow\min\{D^{\prime}{}_{i,j}\mid P_{l}^{i,j}=1\}/r_{c}$
\State $D^{\prime}\leftarrow D^{\prime}-r_{c}t_{l}P_{l}$ \State
$E^{\prime}\leftarrow D^{\prime}$ \State $T^{\prime}\leftarrow T^{\prime}+t_{l}+\delta$
\State$l\leftarrow l+1$ \Else \State$\gamma\leftarrow\gamma/2$
\EndIf\EndWhile\State $L\leftarrow l-1$ \State $E,T\leftarrow$\textit{Migrating}
$\left(E^{\prime},T^{\prime}\right)$\end{algorithmic} \label{alg:alg2}
}
\end{algorithm}

Algorithm \ref{alg:alg2} involves four primary operations: \textit{Regularization}
\cite{regularization}, \textit{Stuffing},\textit{ Slicing }and\textit{
Migrating. Regularization }(line 1) is a simple but efficient pre-processing
operation that can significantly reduce the frequency of reconfigurations
while minimizing the impact on circuit idle time \cite{regularization}.\textit{
Stuffing }(line 1) entails adding artificial demands to the original
demand matrix $D$ to create a $K$-bistochastic demand matrix $D^{\prime}$,
so that it can be decomposed by the BvN theory.\textit{ Slicing }(lines
5-16) is based on BvN and exploits the decomposability of $K$-bistochastic
matrices to iteratively compute a long-duration scheduling plan, greedily
avoiding short and inefficient configurations. Note that the residual
demand $E^{\prime}$ generated by the current iteration is the input
matrix $D^{\prime}$ for the next iteration.\textit{ Slicing} terminates
when the residual demand matrix $E^{\prime}$ (a bistochastic matrix)
in the current iteration becomes small enough to be accommodated by
the packet switch, i.e., $r_{p}T^{\prime}>\rho_{E^{\prime}}$, where
$\rho_{E^{\prime}}$ is the diameter of the residual matrix $E^{\prime}$
generated by the current iteration. Finally, since $r_{p}T^{\prime}>\rho_{E^{\prime}}$,
this means that the current residual matrix $E^{\prime}$ requires
less time to be transmitted through the packet switch than the total
completion time $T^{\prime}$ rather than strictly equal, allowing
for further optimization. Therefore, we propose \textit{Migrating}
(line 18), a new operation that allocates more traffic load to the
packet switch to generate the final residual matrix $E$ ($\rho_{E}>\rho_{E^{\prime}}$)
and the final total completion time $T$ ($T<T^{\prime}$), ensuring
that both switches complete their transmissions simultaneously (i.e.,
$r_{p}T=\rho_{E}$) and further optimizing overall performance.

\subsubsection{Regularization}

In practical applications, if reconfiguration delays in OCS are not
negligible, coflow scheduling based on the BvN decomposition may lead
to poor CCT. This is since the original BvN-based coflow scheduling
often requires preemption, leading to frequent reconfigurations, whereas
non-preemptive scheduling may result in long circuit idle time. \textit{Regularization}
\cite{regularization} is a pre-processing technique that can be used
to reduce the frequency of reconfiguration at a low cost in terms
of circuit idle time. It adjusts each element $d_{i,j}\in D$ to $\left\lceil \frac{d_{i,j}}{\delta}\right\rceil \cdot\delta$,
which is an integer multiple of $\delta$ (i.e., the reconfiguration
delay) to obtain a new regularized matrix $D^{*}$. Since each element
of the new matrix is larger than the original $D$, an efficient scheduling
solution that satisfies the new matrix will also satisfy the original
demand. We adopt the existing operation of \textit{Regularization}
\cite{regularization} to handle a demand matrix of coflow, which
leads to a considerably less frequent circuit reconfiguration.

\subsubsection{Stuffing}

\textit{Stuffing} is the process of converting a regularized matrix
$D^{*}$ into a $K$-bistochastic matrix $D^{\prime}$ by adding artificial
demands. QuickStuff \cite{solstice} is used to perform this operation,
which stuffs the non-zero elements of $D^{\prime}$ in any order.
Then, it checks the zero elements and adds them if necessary until
a $K$-bistochastic matrix is obtained. In our work, \textit{Stuffing}
does not increase the maximum row/column sum (i.e., $\rho$) or the
maximum number of non-zero elements of each row or column (i.e., $\tau$),
which means that $\rho_{D^{\prime}}=\rho_{D^{*}}$ and $\tau_{D^{\prime}}=\tau_{D^{*}}$.

\subsubsection{Slicing}

After \textit{Stuffing}, our algorithm enters its third phase, \textit{Slicing},
which is logically equivalent to the primary iteration of BvN. We
must iteratively determine the next circuit configuration and the
corresponding duration. However, there is no known algorithm that
can explore all possible configurations in polynomial time, so we
employ a greedy approach. Our method, in contrast to BvN, selects
configurations with longer durations to compensate for the reconfiguration
cost and maintain higher utilization rates. Moveover, unlike BvN,
the \textit{Slicing} process terminates once the packet switch is
capable of forwarding the residual traffic demand.

In each iteration of \textit{Slicing}, we input the current demand
matrix $D^{\prime}$, which is the residual matrix $E^{\prime}$ resulting
from the previous iteration, as well as a threshold $\gamma$, and
obtain a circuit configuration $P_{l}$ as the output. To determine
the circuit configuration, we regard the demand matrix $D^{\prime}$
as a bipartite graph between the senders and the receivers and search
for a perfect matching of size $N$ with the largest minimum element,
which is known as Maximum Weighted Minimum Matching (MMWM) \cite{solstice}.
The minimum element of each circuit configuration determines its duration,
and we start with a high threshold value $\gamma$, which is the largest
power of 2 less than the largest element in $D^{\prime}$. We attempt
to find a perfect matching on the demand matrix, ignoring the values
below the threshold, so that any perfect matching returned has a duration
of at least $\gamma/r_{c}$ \cite{solstice}. We repeat the process
with the same threshold until no more perfect matchings can be found
at that threshold, and then reduce the threshold by half before the
next iteration.

The \textit{Slicing} operation ends when the packet switch has sufficient
capacity to handle the residual demand, which is tracked by matrix
$E^{\prime}$. In the next iteration, the previous $E^{\prime}$ becomes
the current demand matrix $D^{\prime}$. The total time required to
schedule the traffic demand is recorded by variable $T^{\prime}$,
which includes transmission time $T_{trans}^{\prime}$ and reconfiguration
delay time $T_{comf}^{\prime}$ (i.e., $T^{\prime}=T_{trans}^{\prime}+T_{conf}^{\prime}$).
Once the time $T^{\prime}$ is large enough to allow the packet switch
to handle the residual demand $E^{\prime}$, i.e., $r_{p}T^{\prime}>\rho_{E^{\prime}}$,
\textit{Slicing} terminates.

\subsubsection{Migrating}

Let $T^{\prime}$ and $T$ represent the total time at the end of
\textit{Slicing }(i.e., before\textit{ Migrating}) and the final total
time after\textit{ Migrating}, respectively. When the condition $T_{E^{\prime}}^{\prime}=\frac{\rho_{E^{\prime}}}{r_{p}}<T^{\prime}$
(i.e., $r_{p}T^{\prime}>\rho_{E^{\prime}}$) is satisfied, where $T_{E^{\prime}}^{\prime}$
denotes the current completion time of the residual matrix $E^{\prime}$,
indicating that the packet switch completes the transmission early,
resulting in a low link utilization for the packet switch due to the
unused of time slots of $T^{\prime}-T_{E^{\prime}}^{\prime}$. To
maximize link utilization over the packet switch, we can redistribute
some of the load from the circuit switch to the packet switch, ensuring
they both remain active for the same duration (i.e., completing the
transmission simultaneously), thus further reducing the total completion
time.

Hence, when\textit{ Slicing} ends, a novel and effective operation,\textit{
}called \textit{Migrating}, is performed. Specifically, we select
a circuit configuration matrix $P_{\tilde{l}}$, and based on the
position of the port connections in $P_{\tilde{l}}$ (i.e., $P_{\tilde{l}}^{i,j}=1$
), by migrating partial data $\varepsilon$ into the current residual
matrix $E^{\prime}$ at the same position, obtain the final residual
matrix $E$ $(\tau_{E}\leq\tau_{E^{\prime}}+1)$, so that the transmission
time of $E$ (i.e., $\frac{\rho_{E}}{r_{p}}$, where $\rho_{E}>\rho_{E^{\prime}}$)
on the packet switch is equal to the final total time $T$ after \textit{Migrating}.
In other words, both the circuit switch and the packet switch complete
the data transmission simultaneously (i,e., $r_{p}T=\rho_{E}$, $T<T^{\prime}$).
According to $r_{p}T=\rho_{E}$, we can calculate the $\varepsilon$
through

\begin{equation}
\begin{aligned}r_{p}\left(T_{trans}+T_{conf}\right)=r_{p}\left[\left(\sum_{l=1,l\neq\tilde{l}}^{L}t_{l}+\left(t_{\tilde{l}}-\frac{\varepsilon}{r_{c}}\right)\right)+L\delta\right]\\
=r_{p}\left[\left(T_{trans}^{\prime}-\frac{\varepsilon}{r_{c}}\right)+T_{conf}^{\prime}\right]=\rho_{E^{\prime}}+\varepsilon
\end{aligned}
,\label{eq:Migrating}
\end{equation}
where, $T_{trans}+T_{conf}=T$ and $\rho_{E^{\prime}}+\varepsilon=\rho_{E}$.
The duration of the selected configuration ($t_{\tilde{l}}$) is reduced
(i.e., $T_{trans}<T_{trans}^{\prime}$), while the number of configurations
remains the same (i.e., $T_{conf}^{\prime}=T_{conf}$). Consequently,
the final total completion time is further reduced (i.e., $T<T^{\prime}$),
and the system performance is optimized. In fact, we can randomly
select a $P_{\tilde{l}}$ as long as the corresponding duration satisfies
$r_{c}t_{\tilde{l}}>\varepsilon$.

\subsubsection{Example}

\begin{figure*}[t]
\centering\includegraphics[width=0.62\textwidth,height=12cm]{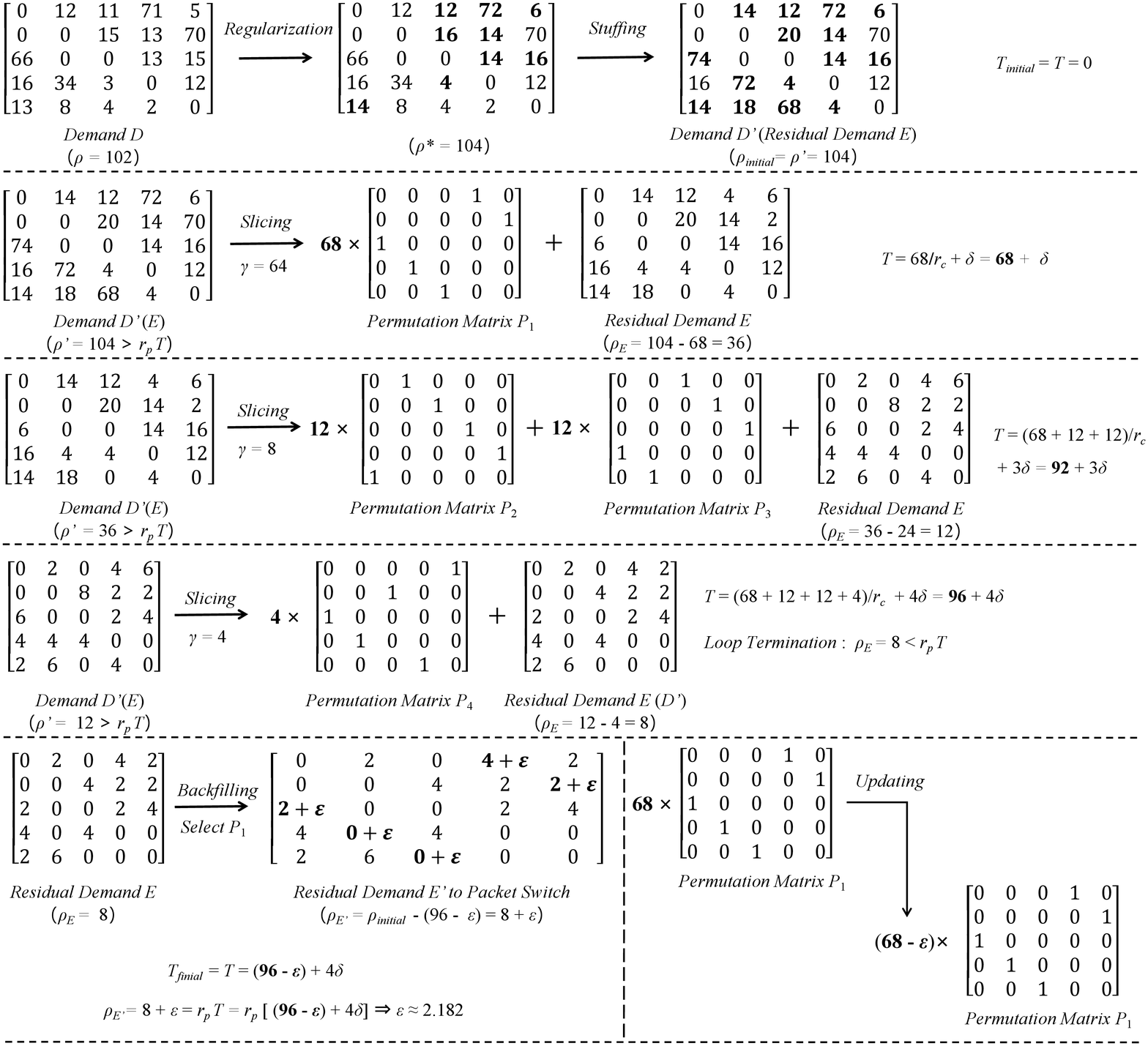}
\caption{An Example Execution}
\label{fig:example}
\end{figure*}

Consider the example depicted in Fig. \ref{fig:example} to illustrate
how our algorithm operates, assuming a fixed time of delay $\delta$
= 2, and packet and circuit rates of $r_{p}$ = 0.1 and $r_{c}$ =
1, respectively. We define the diameter of a matrix as the maximum
row or column sum, denoted by $\rho$. The diameter of the input demand
matrix $D$ is 102 (i.e., the fifth column sum), and $D$ is regularized
to yield the matrix $D^{*}$ with $\rho^{*}$ = 104. Next, we perform
\textit{Stuffing} on $D^{*}$ to obtain a new matrix $D^{\prime}$
, where the sum of each row and column is also 104 (i.e., $D^{\prime}$
is $K$-bistochastic with $K$ = 104). The diameter before and after
\textit{Stuffing} needs to remain constant, i.e., $\rho^{*}=\rho^{\prime}$.

During the first iteration of the algorithm, we choose $\gamma$ =
64 and search for a subset of elements with values at least 64. We
only find one perfect matching with a minimum value of 68, so the
duration of the first configuration is $68/r_{c}$. We then obtain
the current residual matrix $E^{\prime}$ by subtracting the demand
from the current matrix $D^{\prime}$. The current total time is then
updated to $T^{\prime}=68/r_{c}+1\delta=70$. If $r_{p}T$ (in this
case, 7.0) is greater than the diameter of $D^{\prime}$ ($\rho^{\prime}=36$),
then we can transfer the residual demand directly to the packet switch.
Nonetheless, this condition is not satisfied, so we continue to perform
\textit{Slicing} with decreasing thresholds. Since our algorithm seeks
for perfect matchings, it needs to consider at least $N$ elements
and ensure that the found $P_{m}$ is a non-singular permutation matrix
(i.e., rank is $N$), rather than a sub-permutation matrix (i.e.,
rank is less than $N$). When $\gamma$ = 32 and $\gamma$ = 16, perfect
matchings cannot be obtained, therefore the threshold is reduced once
again. When $\gamma$ = 8, there are two perfect matchings for $D^{\prime}$with
the same minimum element of size 12. The total time is now $T^{\prime}=(68+12+12)/r_{c}+3\delta=98$,
and $r_{p}T^{\prime}$ (here 9.8) is smaller than the diameter of
$D^{\prime}$ ($\rho^{\prime}=12$), so the loop continues. The loop
(\textit{Slicing}) ends when the current total time is $T^{\prime}=(68+12+12+4)/r_{c}+4\delta=104$,
and $r_{p}T^{\prime}>\rho_{E^{\prime}}$ (here $10.4>8$). To further
optimize performance, we perform \textit{Migrating }and select a configuration
matrix $P_{1}$ at random, by migrating $\varepsilon$ (calculated
by Eq (\ref{eq:Migrating})) data into the current residual matrix
$E^{\prime}$, obtaining the final residual matrix $E$ such that
$r_{p}T=\rho_{E}$ and $T<T^{\prime}$. Finally, we need to update
the duration of the configuration $P_{1}$. In fact, we may also select
$P_{2}$, $P_{3}$ or $P_{4},$ given that their durations satisfy
$r_{c}t_{l}>\varepsilon$.

\section{Theoretical Analysis}

In this section, we prove our proposed algorithm is $O(\tau)$-approximate,
where $\tau$ is a factor related to demand characteristics, which
is the first approximation algorithm for single coflow scheduling
in hybrid networks.

\textbf{Circuit-Switched Lower Bound:} For a circuit switch, to satisfy
the traffic demand $D$, the total transmission time, $T_{trans}$,
should be at least as much as the largest row or column sum, diameter
$\rho$, divided by the link rate $r_{c}$. Furthermore, since the
circuit switch needs to be configured at least as many times as the
maximum number of non-zero elements of each row or column, denoted
as $\tau$, and each configuration incurs a penalty of $\delta$,
the total configuration time $T_{conf}$ is at least $\tau\delta$.
Therefore, in a pure circuit-switched network, we can obtain the following
lower bound:

\begin{equation}
T_{LB}^{C}=\frac{\rho}{r_{c}}+\tau\delta.\label{eq:T_LB}
\end{equation}

\textbf{Hybrid-Switched Lower Bound:} In a hybrid switch, because
it is possible to divert (small-volume) data via a packet switch,
the number of needed configurations may be reduced by relaxing the
value of $\tau$ from the count of non-zero elements of each row and
column to either 1 or 0 \cite{solstice}. As a result, we reduce the
total circuit switch time from Eq. (\ref{eq:T_LB}) to Eq. (\ref{eq:T_LB_reduced})
, which is proportional to the ratio of circuit link bandwidth $r_{c}$
to total bandwidth due to the introduction of the packet switch:

\begin{equation}
T_{LB}^{H}=\left(\frac{r_{c}}{r_{c}+r_{p}}\right)\left(\frac{\rho}{r_{c}}+\delta\right).\label{eq:T_LB_reduced}
\end{equation}

When we analyze the performance of the scheduling algorithm, we need
to consider the theoretical lower bounds of CCT, i.e., $T_{LB}^{C}$
and $T_{LB}^{H}$, that indicate the optimal theoretical limits of
CCT independent of the scheduling algorithms. In practice, the achievable
CCT may be much larger than the lower bound. Nevertheless, we can
still evaluate the algorithm's efficiency by comparing its performance
with these lower bounds.

\begin{lemma} Reco-Sin is 2-approximate, i.e., $T^{Reco-Sin}<2\left(\frac{\rho}{r_{c}}+\tau\delta\right)\leq2T^{*}$.
\end{lemma}
\begin{IEEEproof}
Reco-Sin \cite{regularization} is an efficient 2-approximation algorithm
for coflow scheduling in circuit-switched networks. It applies \textit{Regularization}
and \textit{Stuffing} on $D$ with $\delta$ to obtain a new matrix
$D^{\prime}$. $L$ is the number of configurations obtained from
Reco-Sin, the transmission time is $T_{trans}^{Reco-Sin}=\sum_{l=1}^{L}t_{l}$,
and the configuration time is $T_{conf}^{Reco-Sin}=L\delta$. Each
element $d_{i,j}\in D$ is regularized to $d_{ij}^{\prime}=\left\lceil \frac{d_{i,j}}{\delta}\right\rceil \cdot\delta$,
which is an integer multiple of $\delta$. Therefore, the duration
of each circuit configuration is at least $\delta$, ensuring that
$t_{l}\geq\delta$ and therefore $T_{trans}^{Reco-Sin}\geq T_{conf}^{Reco-Sin}$. 

Let $\rho$ and $\rho^{\prime}$ denote the maximum value of the sum
of each row and column of $D$ and $D^{\prime}$, respectively, and
let $\tau$ represent the maximum number of non-zero elements of each
row or column of $D$. \textit{Regularization} increases each element
in $D$ by no more than $\delta$, so $\rho^{\prime}<\rho+\tau\delta$.
In addition, we have $\frac{\rho^{\prime}}{r_{c}}=T_{trans}^{Reco-Sin}$.
Hence, we can derive that
\[
\begin{aligned}T^{Reco-Sin}=T_{trans}^{Reco-Sin}+T_{conf}^{Reco-Sin}\leq2T_{trans}^{Reco-Sin}=2\frac{\rho^{\prime}}{r_{c}}\\
<2\left(\frac{\rho+\tau\delta}{r_{c}}\right)<2\left(\frac{\rho}{r_{c}}+\tau\delta\right)\leq2T^{*},
\end{aligned}
\]
where $T^{Reco-Sin}$ is the CCT given by Reco-Sin \cite{regularization},
and $T^{*}$ is the optimal CCT in a circuit-switched network.

\noindent This completes the proof. Details can be seen in Reco-Sin
\cite{regularization}.
\end{IEEEproof}
\begin{lemma} $T^{Reco-Sin}\geq\left(1+\frac{r_{p}}{r_{c}}\right)T^{Ours}+\left(\tau_{E}-1\right)\delta$,
which is the CCT bound between Reco-Sin and our proposed algorithm.\end{lemma}
\begin{IEEEproof}
Let $E^{\prime}$ and $E$ represent the residual matrix at the end
of \textit{Slicing} (i.e., before\textit{ Migrating}) and the actual
residual matrix to be transmitted to the packet switch after\textit{
Migrating}, respectively. Let $\rho_{E^{\prime}}$ and $\rho_{E}$
represent the maximum value of the sum of each row and column of $E^{\prime}$
and $E$, respectively. Let $\tau_{E^{\prime}}$ and $\tau_{E}$ be
the maximum number of non-zero elements of each row or column of $E^{\prime}$
and $E$, respectively. Assume $T_{E^{\prime}}^{Reco-Sin}$ and $T_{E}^{Reco-Sin}$
are the completion times of $E^{\prime}$ and $E$ based on the algorithm
Reco-Sin \cite{regularization}, respectively.

In reality, the performance improvement of our proposed algorithm
over Reco-Sin is the result of two factors: (1) allowing the simultaneous
transmission of the residual matrix on the packet switch, thereby
reducing the total completion time; (2) performing the \textit{Migrating}
operation, and which increases the load on the packet switch, thereby
further reducing the total completion time.

For a given coflow with demand matrix $D$, let $T^{Reco-Sin}$, $T^{Ours}$
be respectively the CCT given by Reco-Sin \cite{regularization} and
by our proposed algorithm. If we do not perform the \textit{Migrating}
operation, then the time saved by our algorithm compared to Reco-Sin
is $T_{E^{\prime}}^{Reco-Sin}$, i.e., $T^{Reco-Sin}-T^{Ours}=T^{Saved}=T_{E^{\prime}}^{Reco-Sin}$.
According to $T_{LB}^{C}$ (Eq. (\ref{eq:T_LB})), we can get $T^{Saved}\geq\frac{\rho_{E^{\prime}}}{r_{c}}+\tau_{E^{\prime}}\delta$.
However, after \textit{Migrating}, the current $T_{E}^{Reco-Sin}$
is not equal to the current $T^{Saved}$. Due to \textit{Migrating}
operation, we have $\rho_{E}=r_{p}T^{Ours}$ and $\tau_{E}\leq\tau_{E^{\prime}}+1$.
Consequently,

\[
\begin{aligned}T^{Reco-Sin}-T^{Ours}=T^{Saved}\geq\frac{\rho_{E}}{r_{c}}+\tau_{E^{\prime}}\delta\\
\geq\frac{\rho_{E}}{r_{c}}+\left(\tau_{E}-1\right)\delta=\frac{r_{p}T^{Ours}}{r_{c}}+\left(\tau_{E}-1\right)\delta.
\end{aligned}
\]
Recall that $T^{Reco-Sin}-T^{Saved}=T^{Ours}$. We have

\[
\begin{aligned}T^{Saved}\geq\frac{r_{p}T^{Ours}}{r_{c}}+\left(\tau_{E}-1\right)\delta\\
\geq\frac{r_{p}}{r_{c}}\left(T^{Reco-Sin}-T^{Saved}\right)+\left(\tau_{E}-1\right)\delta,
\end{aligned}
\]
thus,

\[
\left(1+\frac{r_{p}}{r_{c}}\right)T^{Saved}\geq\frac{r_{p}}{r_{c}}T^{Reco-Sin}+\left(\tau_{E}-1\right)\delta.
\]
Next, we have

\[
\left(1+\frac{r_{p}}{r_{c}}\right)\left(T^{Reco-Sin}-T^{Ours}\right)\geq\frac{r_{p}}{r_{c}}T^{Reco-Sin}+\left(\tau_{E}-1\right)\delta.
\]
Finally,

\[
T^{Reco-Sin}\geq\left(1+\frac{r_{p}}{r_{c}}\right)T^{Ours}+\left(\tau_{E}-1\right)\delta.
\]
This completes the proof.
\end{IEEEproof}
Set $T^{Ours}$ and $T^{*}$ are respectively the CCT of our algorithm
and optimal CCT in a hybrid-switched network. Based on the above Lemma
1 and 2, we can get Theorem 1.
\begin{thm}
Our proposed single coflow scheduling algorithm is $O(\tau)$-approximate,
i.e., $T^{Ours}\leq O(\tau)T^{*}$.
\end{thm}
\begin{IEEEproof}
For a hybrid-switched network, because $T_{LB}^{H}$ (Eq. (\ref{eq:T_LB_reduced}))
is a lower bound of CCT for any algorithm, we have $T^{*}\geq T_{LB}^{H}=\left(\frac{r_{c}}{r_{c}+r_{p}}\right)\left(\frac{\rho}{r_{c}}+\delta\right)$.

\noindent By Lemma 1, we can get $T^{Reco-Sin}<2\left(\frac{\rho}{r_{c}}+\tau\delta\right)$.
Hence,

\[
\begin{aligned}\left(\frac{r_{c}}{r_{c}+r_{p}}\right)T^{Reco-Sin}<2\left(\frac{r_{c}}{r_{c}+r_{p}}\right)\left(\frac{\rho}{r_{c}}+\tau\delta\right)\\
=2\left(\frac{r_{c}}{r_{c}+r_{p}}\right)\left(\frac{\rho}{r_{c}}+\delta\right)+2\left(\frac{r_{c}}{r_{c}+r_{p}}\right)\left(\tau-1\right)\delta\\
\leq2T^{*}+2\left(\frac{r_{c}}{r_{c}+r_{p}}\right)\left(\tau-1\right)\delta\text{.}
\end{aligned}
\]
Further, by Lemma 2, we can get

\[
\begin{aligned}\left(\frac{r_{c}}{r_{c}+r_{p}}\right)\left[\left(1+\frac{r_{p}}{r_{c}}\right)T^{Ours}+\left(\tau_{E}-1\right)\delta\right]\\
\leq\left(\frac{r_{c}}{r_{c}+r_{p}}\right)T^{Reco-Sin}\leq2T^{*}+2\left(\frac{r_{c}}{r_{c}+r_{p}}\right)\left(\tau-1\right)\delta.
\end{aligned}
\]
Since $1\leq\tau_{E}\leq N$, hence,

\[
\begin{aligned}T^{Ours}\leq2T^{*}+\left(\frac{r_{c}}{r_{c}+r_{p}}\right)\left(2\tau-\tau_{E}-1\right)\delta\\
\leq2T^{*}+\left(\frac{r_{c}}{r_{c}+r_{p}}\right)\left(2\tau-2\right)\delta.
\end{aligned}
\]
Recall that $T^{*}\geq\left(\frac{r_{c}}{r_{c}+r_{p}}\right)\left(\frac{\rho}{r_{c}}+\delta\right)$,
thus,

\[
\begin{aligned}\frac{\left(\frac{r_{c}}{r_{c}+r_{p}}\right)\left(2\tau-2\right)\delta}{T^{*}}\leq\frac{\left(\frac{r_{c}}{r_{c}+r_{p}}\right)\left(2\tau-2\right)\delta}{\left(\frac{r_{c}}{r_{c}+r_{p}}\right)\left(\frac{\rho}{r_{c}}+\delta\right)}\\
=\frac{\left(2\tau-2\right)\delta}{\left(\frac{\rho}{r_{c}}+\delta\right)}<\frac{\left(2\tau-2\right)\delta}{\delta}=2\tau-2.
\end{aligned}
\]
Thus we have

\[
\left(\frac{r_{c}}{r_{c}+r_{p}}\right)\left(2\tau-2\right)\delta\leq\left(2\tau-2\right)T^{*}.
\]
Finally,

\[
T^{Ours}\leq2T^{*}+\left(2\tau-2\right)T^{*}=2\tau T^{*}=O(\tau)T^{*},
\]
where $\tau$ is the demand characteristic, i.e., the maximum number
of non-zero elements of each row or column of the input matrix $D$.

\noindent This completes the proof.
\end{IEEEproof}

\section{Experimental Evaluations}

In this section, we use the traces of Facebook \cite{facebook} to
test the performance of the proposed method and provide simulation
results and detailed performance analysis.

\subsection{Simulation Settings}

\textbf{Workload: }Our workload is generated based on Facebook trace
\cite{facebook}, which is collected from a 3000-machine, 150-rack
MapReduce cluster at Facebook. This trajectory is extensively used
in simulation \cite{sunflow,regularization}, it contains 526 coflows
scaled down to a 150-port fabric with exact inter-arrival times. For
each coflow in the Facebook trace, the sender machines, receiver machines,
and the transmitted bytes at the receiver level, rather than the flow
level, are provided. To generate flows, we therefore pseudo-uniformly
divide the bytes from each receiver to each sender. We randomly select
$N$ machines from the trace as servers.

\textbf{Evaluation Metrics: }We evaluate schemes based on the Normalized
Reconfiguration Frequency (Normalized RF) and the Normalized CCT.

$\bullet$ \textit{Normalized RF} is defined as the number of configurations
under the compared scheduler normalized by our algorithm\textquoteright s
RF, i.e.,

\[
\textrm{Norm.RF}=\frac{\textrm{Compared RF}}{\textrm{RF under our algorithm}}.
\]

Obviously, if the \textit{Normalized RF} is greater (smaller) than
one, the benchmark algorithm generates fewer (more) configurations
than the compared scheduler.

$\bullet$ \textit{Normalized CCT} is defined as the CCT under the
compared scheduler normalized by our algorithm\textquoteright s CCT,
i.e.,

\[
\textrm{Norm.CCT}=\frac{\textrm{Compared CCT}}{\textrm{CCT under our algorithm}}.
\]

Intuitively, the benchmark algorithm is faster (slower) if the \textit{Normalized
CCT} is greater (smaller) than one. As a result, this metric can measure
how efficient the benchmark is compared to others.

\textbf{Baseline solutions: }We compare the performances of our proposed
algorithm with the following baselines for single coflow scheduling
in minimizing CCT.

\textit{1)} \textit{Hybrid-Switched Lower Bound}\textsl{: $T_{LB}^{H}=\left(\frac{r_{c}}{r_{c}+r_{p}}\right)\left(\frac{\rho}{r_{c}}+\delta\right)$
}is the optimal theoretical limits of CCT in hybrid-switched networks,
independent of the scheduling algorithms.

\textit{2)} \textsl{Basic BvN \cite{BvN}: }BvN is a fundamental and
classical method for scheduling coflow in optical circuit switches
(OCS), which iteratively calculates the scheduling of configuration
(\textit{Slicing}) to complete the transmission.

\textit{3)} \textit{Reco-Sin \cite{regularization}:}\textsl{ }Reco-Sin
is the first constant approximation algorithm for single coflow scheduling
in OCS. Reco-Sin applies \textit{Regularization} and \textit{Stuffing}
on $D$ with $\delta$, obtaining a doubly stochastic matrix $D^{\prime}$,
and then executes BvN decomposition (\textit{Slicing}) on $D^{\prime}$.

\textit{4)} \textsl{Solstice \cite{solstice}:} Solstice is an efficient
circuit scheduling algorithm in a hybrid network that operates in
two stages: \textit{Stuffing} and \textit{Slicing}.

\subsection{Simulation Results}

Our simulation is based on a hybrid switch with $N$=10 ports. The
hybrid switch consists of a circuit switch with 100 Gbps per link
capacity and a packet switch with 10 Gbps per link capacity. The value
of the reconfiguration delay, $\delta$, ranges from 20 $\mu s$ to
100 $\mu s$, with a default value of 20 $\mu s$.

Essentially, a demand matrix $D$ exhibits sparsity when the proportion
of non-zero elements in the matrix is low. We measure the sparsity
of a matrix with \textit{density}, which is a value between 0 and
1. In this paper, coflows are categorized as sparse, normal, or dense
based on the density of their demand matrix. We consider a matrix
to be sparse when its $density\leq0.2$, normal when $0.2\leq density\leq0.6$,
and dense when $density\geq0.6$. Sparse matrices can be scheduled
more efficiently in circuit switches since they inherently require
fewer configurations \cite{solstice}. 

\begin{figure}[H]
\centering
\centering{}%
\begin{minipage}[c]{0.49\linewidth}%
\centering \includegraphics[width=1\linewidth,height=4.5cm,keepaspectratio]{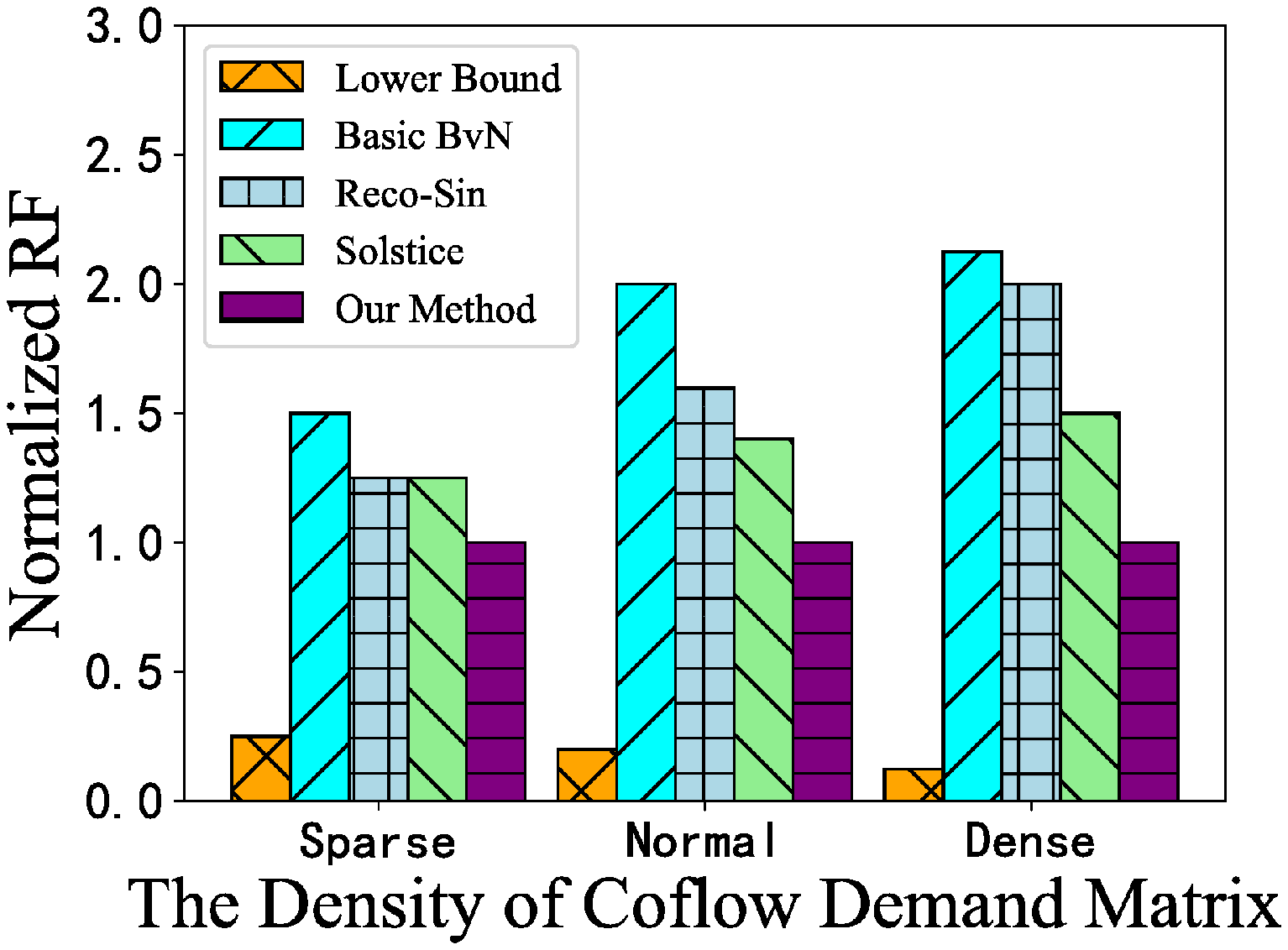}
\caption{Normalized RF in Different Schedulers}
\label{fig:frequency}%
\end{minipage}
\begin{minipage}[c]{0.49\linewidth}%
\centering \includegraphics[width=1\linewidth,height=4.5cm,keepaspectratio]{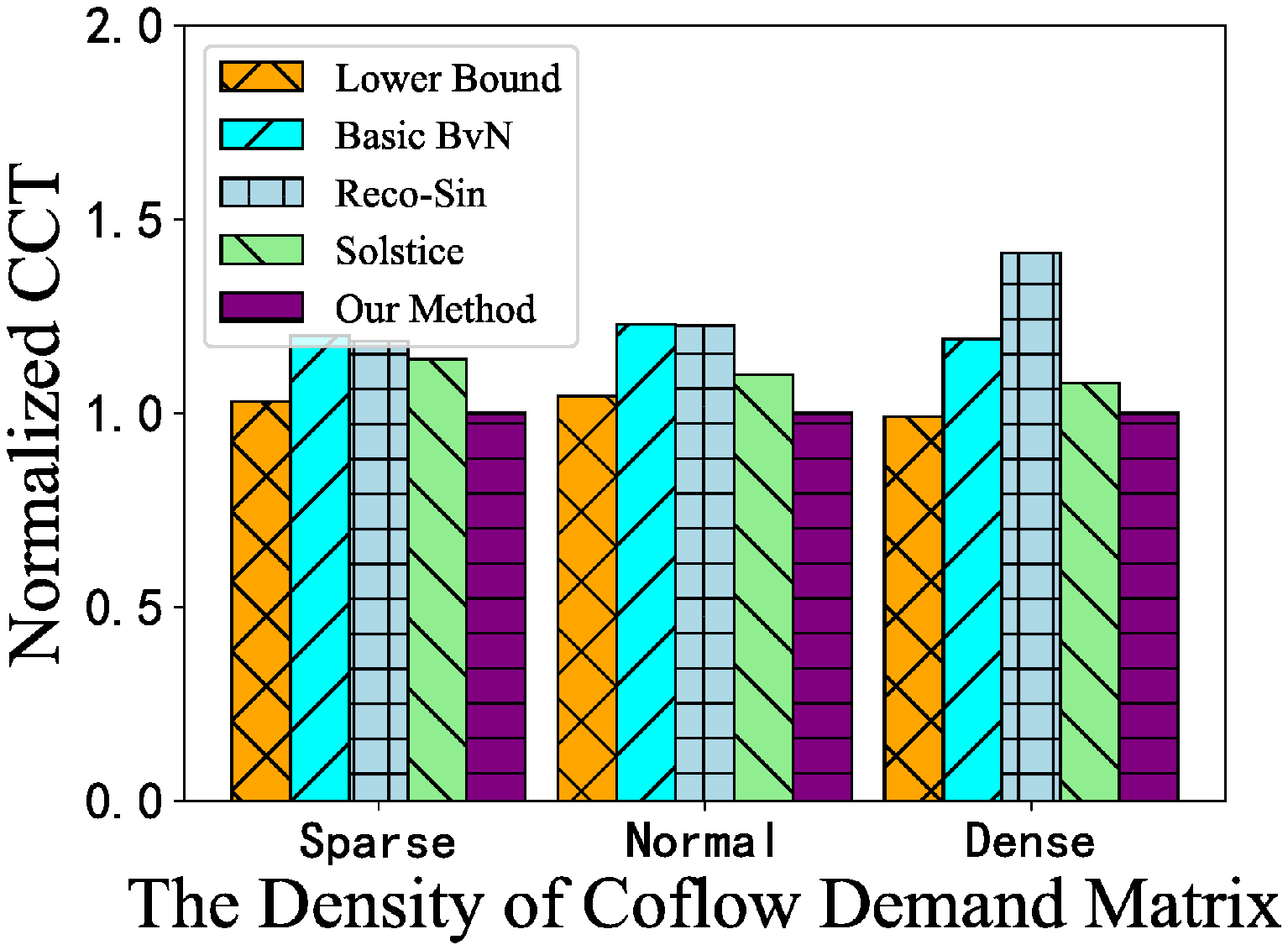}
\caption{Normalized CCT in Different Schedulers}
\label{fig:CCT}%
\end{minipage}
\end{figure}

Fig. \ref{fig:frequency} depicts the reconfiguration frequency of
different algorithms for various density levels, with a fixed reconfiguration
time of 20 $\mu s$. In this case, we utilize the CCT of our algorithm
as a normalized benchmark and show the performance of various schedulers.
The density of the demand matrix can significantly affect the reconfiguration
frequency of the coflow. It is observed that the reconfiguration frequency
of our method is lower than Solstice \cite{solstice}, indicating
that the \textit{Regularization} \cite{regularization} operation
we incorporate to process the traffic matrix indeed results in a significantly
lower circuit reconfiguration frequency. Compared to the Reco-Sin\cite{regularization}
for scheduling coflows in OCS, we utilize the hybrid switched network
to transfer the remaining low-volume traffic to the packet switch,
thereby effectively reducing the reconfiguration frequency.

As shown in Fig. \ref{fig:frequency}, Solstice spends $1.25\times$,
$1.17\times$ and $1.09\times$ more reconfigurations than our method,
when the demand matrix of coflow is sparse, normal and dense, respectively.
Reco-Sin spends $1.25\times$, $1.33\times$ and $1.45\times$ more
reconfigurations than our method, when the demand matrix of coflow
is sparse, normal and dense, respectively. As the density increases,
the performance gap becomes even greater. The reason for this is that
the number of BvN-decomposed permutation matrices in Solstice increases
as the coflow density increases, whereas the role played by \textit{Regularization}
in our method is likely to become increasingly apparent, so the performance
gap increases. In addition, Basic BvN has the highest number of reconstructions
due to the absence of the \textit{Regularization} operation, considering
only OCS and not utilizing EPS.

Next, as shown in Fig. \ref{fig:CCT}, we evaluate the performance
of our method and different schedulers in terms of minimizing the
coflow completion time (CCT). Compared to BvN and Reco-Sin, which
schedule single coflow in optical circuit switches (OCS), our method
allows for the simultaneous transmission of remaining traffic on packet
switches, thereby reducing the total completion time (i.e., CCT).
Furthermore, for Solstice (which also schedules coflows in hybrid
networks), we integrate the \textit{Regularization} process to decrease
reconfiguration frequency, and propose the \textit{Migrating} operation
to further minimize the CCT and optimize system performance, thus
surpassing Solstice in performance. Solstice requires $1.14\times$,
$1.10\times$ and $1.08\times$ more time than our algorithm to schedule
the demand matrix with sparse, normal and dense coflows, respectively.
Additionally, Reco-Sin requires $1.19\times$, $1.23\times$ and $1.42\times$
more time than our algorithm to schedule the demand matrix with sparse,
normal and dense coflows, respectively.

\begin{figure}[H]
\centering
\centering{}%
\begin{minipage}[c]{0.49\linewidth}%
\centering \includegraphics[width=1\linewidth,height=4.5cm,keepaspectratio]{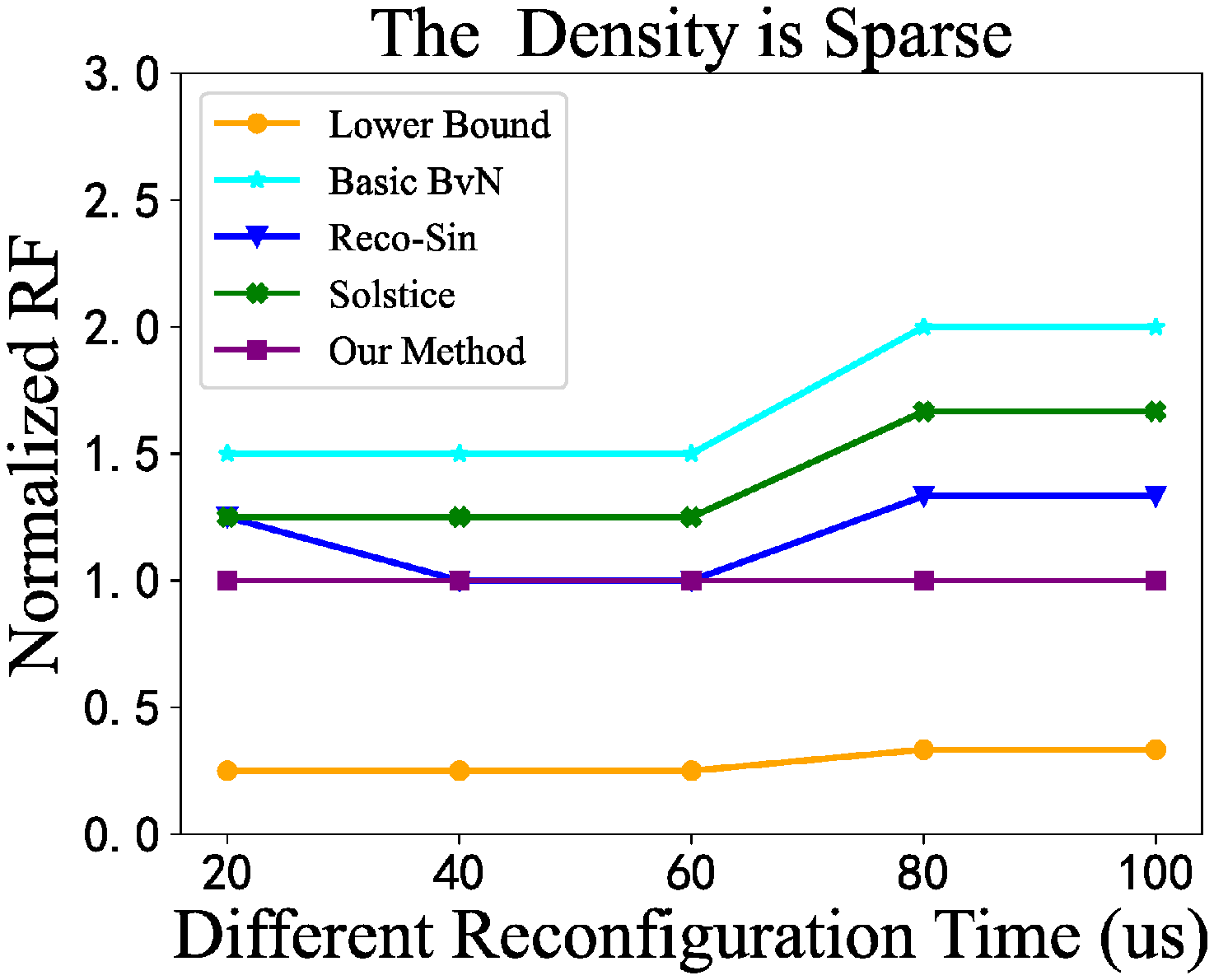}
\caption{Normalized RF in Sparse Density}
\label{fig:rf_spare}%
\end{minipage}
\begin{minipage}[c]{0.49\linewidth}%
\centering \includegraphics[width=1\linewidth,height=4.5cm,keepaspectratio]{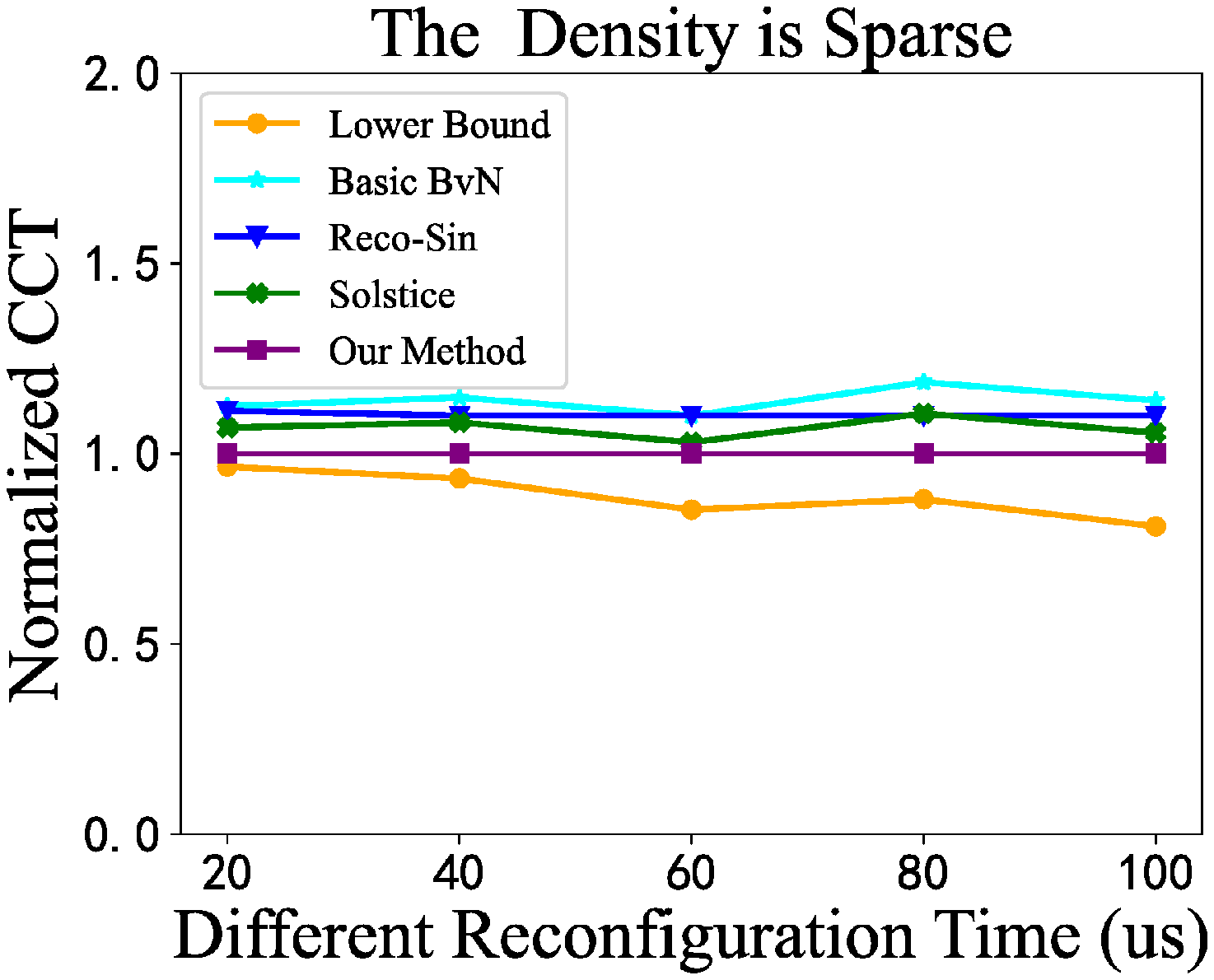}
\caption{Normalized CCT in Sparse Density}
\label{fig:cct_spare}%
\end{minipage}
\end{figure}

\begin{figure}[H]
\centering
\centering{}%
\begin{minipage}[c]{0.49\linewidth}%
\centering \includegraphics[width=1\linewidth,height=4.5cm,keepaspectratio]{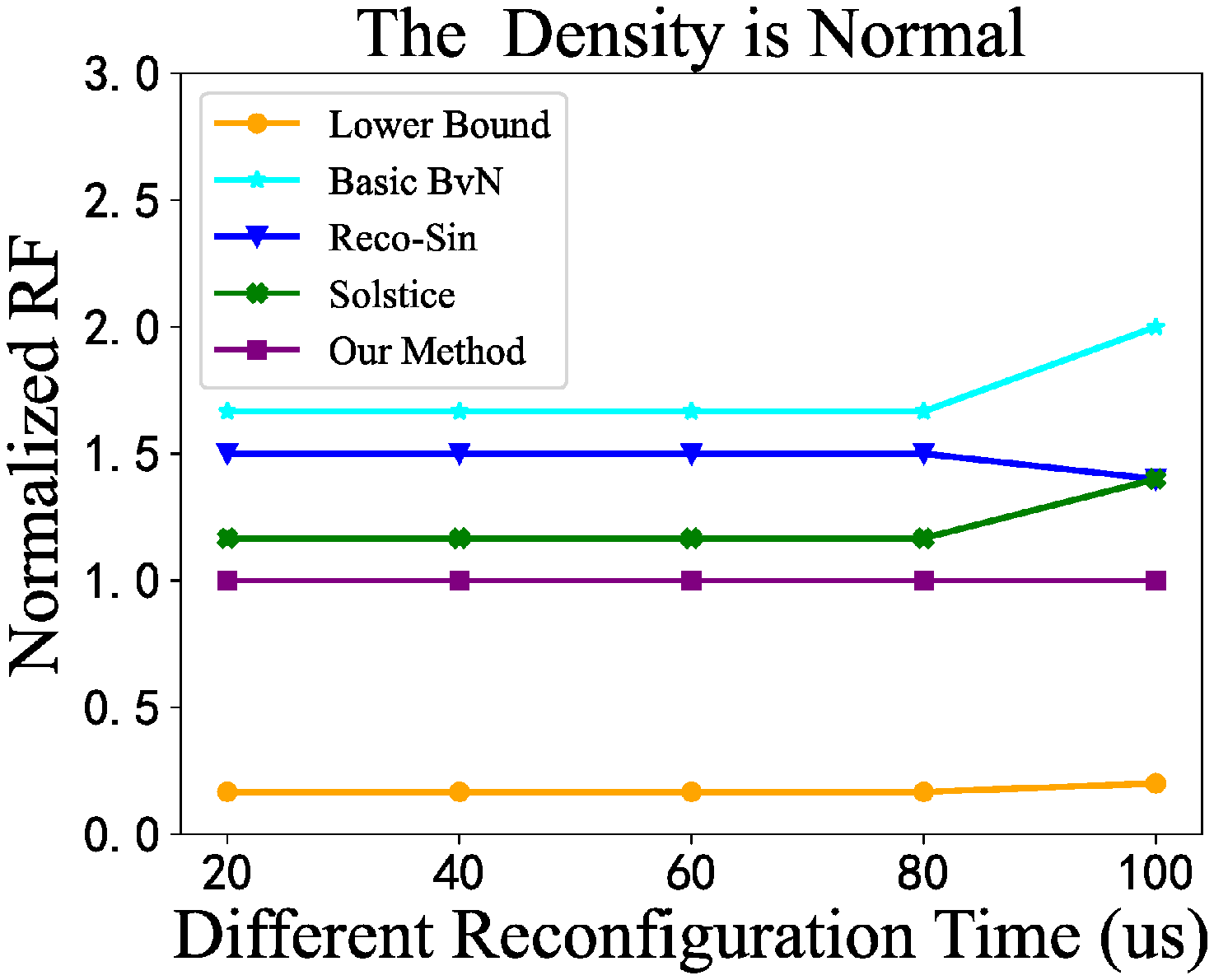}
\caption{Normalized RF in Normal Density}
\label{fig:rf_normal}%
\end{minipage}
\begin{minipage}[c]{0.49\linewidth}%
\centering \includegraphics[width=1\linewidth,height=4.5cm,keepaspectratio]{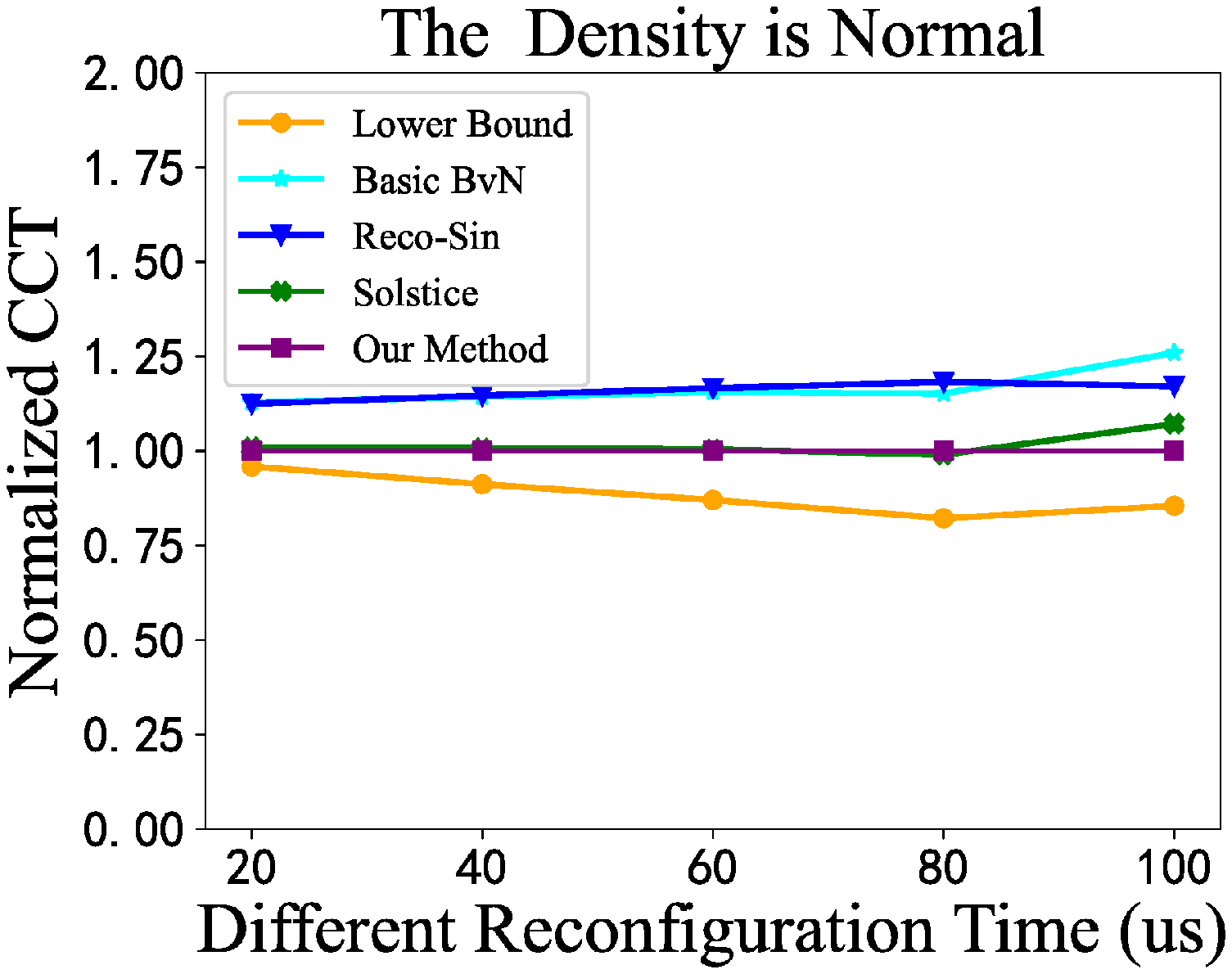}
\caption{Normalized CCT in Normal Density}
\label{fig:cct_normal}%
\end{minipage}
\end{figure}

\begin{figure}[H]
\centering
\centering{}%
\begin{minipage}[c]{0.49\linewidth}%
\centering \includegraphics[width=1\linewidth,height=4.5cm,keepaspectratio]{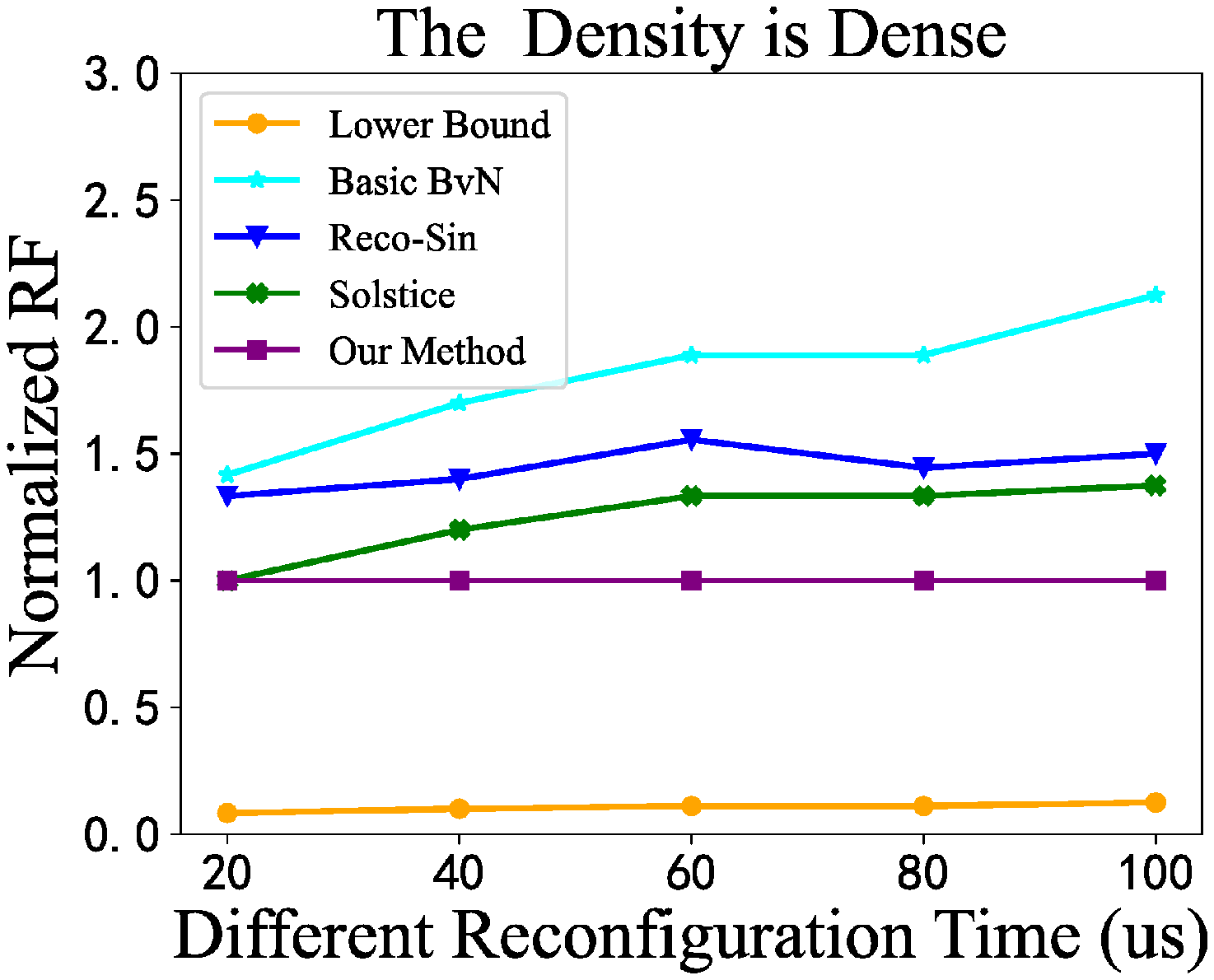}
\caption{Normalized RF in Dense Density}
\label{fig:rf_dense}%
\end{minipage}
\begin{minipage}[c]{0.49\linewidth}%
\centering \includegraphics[width=1\linewidth,height=4.5cm,keepaspectratio]{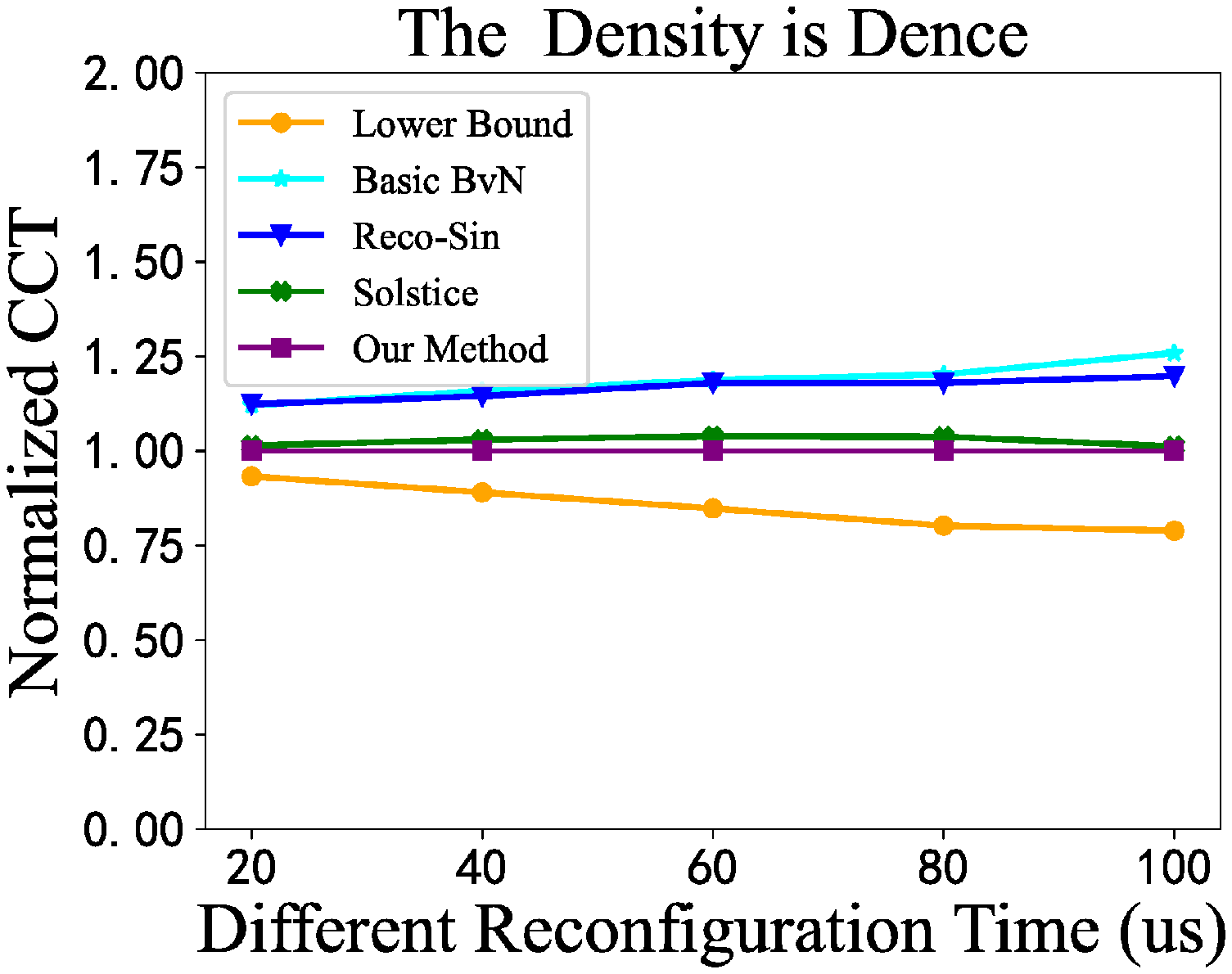}
\caption{Normalized CCT in Dense Density}
\label{fig:cct_dense}%
\end{minipage}
\end{figure}

The variation of $\delta$ is an important property determined by
the hardware of Optical Circuit Switching (OCS), which directly affects
the CCT and indirectly changes the reconfiguration frequency. In this
case, we utilize the theoretical lower bound of RF and CCT as a normalized
benchmark and show the performance of various schedulers. The curves
in Fig. \ref{fig:rf_spare} demonstrate that our proposed algorithm
requires less reconfiguration time (i.e., fewer reconfigurations)
to complete the same coflow compared to Reco-Sin and Solstice. A comparative
analysis of Fig. \ref{fig:rf_spare}, \ref{fig:rf_normal} and \ref{fig:rf_dense}
shows that the number of reconfigurations decreases as $\delta$ increases
for both our method and Reco-Sin. This observation is primarily based
on the fact that the \textit{Regularization} operation is directly
related to $\delta$. As $\delta$ increases, \textit{Regularization}
operation causes the elements of the coflow demand matrix to become
more aligned, thus reducing reconfiguration time. Note that \textit{Regularization}
operation may not be effective in reducing the number of configurations
when the demand matrix is too sparse. Conversely, when $\delta$ varies,
the number of reconfigurations for Solstice remains relatively constant.
In fact, the variation of $\delta$ would have no significant impact
on the reconfiguration frequency of Solstice.

Fig. \ref{fig:cct_spare} displays the performance in terms of CCT.
A comparative analysis of Fig. \ref{fig:cct_spare}, \ref{fig:cct_normal}
and \ref{fig:cct_dense} shows that, as the coflow demand matrix shifts
from sparse to dense, the advantages of our method become more apparent.
This is because as the demand matrix becomes more dense, the required
number of configurations also increases. Furthermore, as $\delta$
increases, reconfiguration time dominates the total completion time
(i.e., CCT). At this point, the advantage of our method becomes more
prominent as it yields fewer reconfigurations.

\section{Conclusions}

In this paper, we explore how to schedule single coflows more efficiently
in hybrid-switched data center networks (DCNs). We incorporate an
existing operation called\textit{ Regularization} \cite{regularization}
to handle a demand matrix of coflow, allowing circuits to be reconfigured
significantly less frequently. We also introduce a new technique called
\textit{Migrating}, which further decreases the CCT and improves system
performance. We then develop an efficient coflow scheduling algorithm
to minimize the CCT and demonstrate that it achieves a performance
guarantee (approximation ratio to the optimal solution) of $O(\tau)$,
where $\tau$ is a factor related to demand characteristics. To the
best of our knowledge, this is the first approximation algorithm for
coflow scheduling in a hybrid-switched DCN. Extensive simulations
based on real data traces show that our proposed algorithm significantly
outperforms the state-of-the-art schemes in terms of reducing the
number of reconfigurations and speeding up the transmission of single
coflow.

\section*{Acknowledgment}

This work is supported by Macao Polytechnic University Research Grant
\#CI401/DEI/2022 and Key-Area Research and Development Plan of Guangdong
Province \#2020B010164003. The corresponding author is Hong Shen.

\bibliographystyle{IEEEtran}
\addcontentsline{toc}{section}{\refname}\bibliography{IEEEabrv,IEEEexample,IEEE_reference}

\vspace{12pt}

\end{document}